\documentclass[a4paper,11pt]{article}
\usepackage{graphicx}
\usepackage{amsfonts}
\usepackage{amssymb}
\begin{document}

\title{Problems with Mannheim's conformal gravity program}
\maketitle
\begin {center}
\author{\large Youngsub Yoon }

\vskip 1cm

{
\it
%\large
%affiliations
Department of Physics and Astronomy\\
Seoul National University, Seoul 151-747, Korea
}
\end{center}

\begin{abstract}
We show that Mannheim's conformal gravity program, whose potential has a term proportional to $1/r$ and another term proportional to $r$, does not reduce to Newtonian gravity at short distances, unless one assumes undesirable singularities of the mass density of the proton. Therefore, despite the claim that it successfully explains galaxy rotation curves, unless one assumes the singularities, it seems to be falsified by numerous Cavendish-type experiments performed at laboratories on Earth whose work have not found any deviations from Newton's theory. Moreover, it can be shown that as long as the total mass of the proton is positive, Mannheim's conformal gravity program leads to negative linear potential, which is problematic from the point of view of fitting galaxy rotation curves, which necessarily requires positive linear potential.

\end{abstract}

\section {Introduction}
Recently, Mannheim's conformal gravity program has attracted much attention as an alternative to dark matter and dark energy \cite{alternatives, case, Newtonian}. However, so far, the only way its validity could be tested was through cosmological considerations. In this paper, we suggest that Mannheim's conformal gravity program seems problematic according to numerous Cavendish-type experiments on Earth. One of our ideas is that Mannheim's conformal gravity program predicts that the gravitational force due to an object depends on its mass distribution even in the case that in which it has a spherical symmetric mass distribution. (i.e., the mass density only depends on the distance from the center of the body.) For example, Newtonian gravity predicts that we can calculate the gravitational force due to the Earth as if all the mass of Earth were at its center. This is not true in the case of conformal gravity; the gravitational force \emph{heavily} depends on the mass distribution.

In Secs. 2 and 3, we introduce and review conformal gravity. In other sections, we give a couple of arguments why conformal gravity is problematic.

\section {Mannheim's conformal gravity}
Instead of Einstein-Hilbert action, in conformal gravity, we have the following action.

\begin{eqnarray}
S=-\alpha_g \int d^4x \sqrt{-g}  C_{\lambda\mu\nu\kappa}C^{\lambda\mu\nu\kappa}\nonumber\\
=-2\alpha_g \int d^4x \sqrt{-g} [R^{\mu\nu} R_{\mu\nu}-\frac{1}{3}R^2]\label{alpha}
\end{eqnarray}
where $C_{\lambda\mu\nu\kappa}$ is the conformal Weyl tensor, and $\alpha_g$ is a purely dimensionless coefficient. By adding this to the action of matter and varying it with respect to the metric, one can obtain the conformal gravity version of the Einstein equation.

\section {The metric solution in conformal gravity}
The following derivation closely follows Mannheim and Kazanas's in Ref. \cite{Newtonian}. [See in particular Eqs. (9), (13), (14) and (16) in their paper.] In case there is a spherical symmetry in the distribution of the mass, one can write the metric as follows:

\begin{equation}
ds^2=-B(r)dt^2+\frac{dr^2}{B(r)}+r^2 d\Omega_2
\end{equation}

Plugging this into the conformal gravity version of the Einstein equation, and assuming that all the matter is inside the radius $r_0$, Mannheim and Kazanas obtain

\begin{equation}
B(r>r_0)=1-\frac{2\beta}{r}+\gamma r
\end{equation}
\begin{equation}
\nabla^4 B(r)=f(r)\label{B}
\end{equation}

The solution is given by
\begin{eqnarray}
2\beta=\frac{1}{6}\int_0^{r_0}dr'r'^4 f(r')\label{beta}\\
\gamma=-\frac{1}{2}\int_0^{r_0}dr'r'^2 f(r')\label{gamma}
\end{eqnarray}
where
\begin{equation}
f(r)\equiv \frac{3}{4\alpha_g B(r)}(T^0_0-T^r_r)\label{f}
\end{equation}
without any approximation whatsoever.
\section{Non-Newtonian potential}
If we ignore $T^r_r$ in the above equation, as it is small, set $T^0_0=\rho$, and use $B(r)\approx 1$, we get:

\begin{eqnarray}
\frac{2\beta}{r}=\frac 1r (\frac{1}{8\alpha_g}\int_0^{r_0}dr'r'^4 \rho)\label{2beta}
\end{eqnarray}

Compare this with the Newtonian case, which is the following:
\begin{eqnarray}
\frac{2\beta}{r}=\frac{2G}{rc^2}\int_0^{r_0}dr'4\pi r'^2 \rho=\frac{2GM_{total}}{rc^2}\label{2betaNewton}
\end{eqnarray}

Thus, unlike in the Newtonian case, we see that in Mannheim's conformal gravity, the gravitational attraction depends not only on the total mass, but also on the mass distribution. Therefore, if two spherically symmetric objects with the same mass but different density distributions yield the same strength of gravitational forces, then conformal gravity is troublesome. On the other hand, if they yield different strengths of gravitational force, in precisely the manner that conformal gravity predicts, then conformal gravity will be verified. Notice that the difference of the gravitational force would be big; it would be in leading order, not in next-to-leading order. For example, if the mass of two objects is the same, but the first one's size is double that of the second one, the former will exert quadruple the amount of gravitational force. Conformal gravity seems troublesome, as many Cavendish-type experiments have been performed, and none of them has detected that gravity depends on the density distribution \cite{Cavendish}. We introduce Mannheim and Kazanas's circumvention of this dilemma in the next section.

\section{The wrong sign of the linear potential term}
Mannheim compares the gravitational potential in Newtonian gravity and conformal gravity in Ref. \cite{alternatives}. He considers the case in which all the matter is inside the region $(r<R)$, and the mass distribution only depends on $r$ (i.e., spherically symmetric). In the case of Newtonian gravity, the potential is given by

\begin{equation}
\nabla^2\phi(\vec{r})=g(\vec{r})
\end{equation}
The solution is given by
\begin{equation}
\phi(\vec{r})=-\frac{1}{4\pi}\int d^3\vec{r'}\frac{g(\vec{r'})}{|\vec{r}-\vec{r'}|}\label{g}
\end{equation}
\begin{equation}
\phi(r>R)=-\frac{1}{r}\int_0^R dr'r'^2 g(r')\label{Newtonr>R}
\end{equation}
\begin{equation}
\phi(r<R)=-\frac{1}{r}\int_0^r dr'r'^2g(r')-\int_r^Rdr'r'g(r')\label{Newtonr<R}
\end{equation}

In the case of Mannheim's conformal gravity, we have
\begin{equation}
\nabla^4 \phi(\vec{r})=h(\vec{r})
\end{equation} 
The solution is given by
\begin{equation}
\phi(\vec{r})=-\frac{1}{8\pi}\int d^3\vec{r'}h(\vec{r'})|\vec{r}-\vec{r'}|\label{h}
\end{equation}
\begin{equation}
\phi(r>R)=-\frac{1}{6r}\int_0^R dr'r'^4h(r')-\frac r2 \int_0^R dr'r'^2h(r')\label{Conformalr>R}
\end{equation}
\begin{equation}
\phi(r<R)=-\frac{1}{6r}\int_0^r dr'r'^4h(r')-\frac 12 \int_r^R dr'r'^3h(r')-\frac r2 \int_0^r dr'r'^2h(r')-\frac{r^2}{6}\int_r^R dr'r'h(r')\label{Conformalr<R}
\end{equation}

Then, Mannheim notes that the following $h(r)$:
\begin{equation}
h(r<R)=-\gamma c^2 \sum_{i=1}^N{\delta(r-r^i)\over r^2}
-{3\beta c^2 \over 2}\sum_{i=1}^N\left[\nabla^2
-{r^2 \over 12}\nabla^4\right]\left[{\delta(r-r_i)\over r^2}\right]\label{hnabla}
\end{equation}
yields the following gravitational potential:

\begin{equation}
\phi (r>R)=-{N\beta c^2 \over r}+{N\gamma c^2 r\over 2}\label{phi}
\end{equation}

Thus, by assuming the singularities of the mass density of the proton, Mannheim tries to circumvent the problem we raised in Sec. 4: we can arbitrarily make $1/r$ potential from a single proton, and if we add them up, they would reproduce Newton's law.

However, a closer look at the last term of Eq. (\ref{Conformalr>R}) shows that this circumvention will not work. Notice that $h$ is the mass density up to a certain positive coefficient. Therefore,

\begin{equation}
\int h r'^2 dr'
\end{equation}
should be equal to the \emph{total} mass of the proton divided by the positive coefficient. This is obvious from the following elementary formula:

\begin{equation}
M=\int 4\pi r^2 \rho dr
\end{equation}

Therefore, the last term of Eq. (\ref{Conformalr>R}) yields the negative linear gravitational potential. However, we know that we want the positive linear gravitational potential to fit the galaxy rotation curve; what we need there is not extra repulsion but extra attraction. Therefore, Mannheim's conformal gravity program seems problematic, unless someone can come up with an argument that $\alpha_g$ in Eq. (\ref{alpha}) can take a negative value.

Finally, we want to note that Mannheim's conformal gravity program, with which we have dealt in this paper, should not be confused with Anderson-Barbour-Foster-Murchadha conformal gravity \cite{Anderson}.

\end{document}